\documentclass[aps,nofootinbib,prd,pacs]{revtex4} 
\usepackage{amsfonts}
\usepackage{amssymb}
\usepackage{graphicx}
\usepackage{pstricks}
\usepackage{epsfig}

\newcommand{\be}{\begin{equation}}
\newcommand{\ee}{\end{equation}}
\newcommand{\beq}{\begin{eqnarray}}
\newcommand{\eeq}{\end{eqnarray}}

\def\lsim{\hbox{ \raise.35ex\rlap{$<$}\lower.6ex\hbox{$\sim$}\ }}
\def\gsim{\hbox{ \raise.35ex\rlap{$>$}\lower.6ex\hbox{$\sim$}\ }} 

\begin{document}
\title{Lattice Refining Loop Quantum Cosmology and Inflation}
\author{William Nelson and Mairi Sakellariadou}
\affiliation{King's College London,
Department of Physics, Strand WC2R 2LS, London, U.K.}

\begin{abstract}
\vspace{.2cm}
\noindent
We study the importance of lattice refinement in achieving a
successful inflationary era. We solve, in the continuum limit, the
second order difference equation governing the quantum evolution in
loop quantum cosmology, assuming both a fixed and a dynamically
varying lattice in a suitable refinement model. We thus impose a
constraint on the potential of a scalar field, so that the continuum
approximation is not broken.  Considering that such a scalar field
could play the r\^ole of the inflaton, we obtain a second constraint
on the inflationary potential so that there is consistency with the
CMB data on large angular scales. For a $m^2\phi^2/2$ inflationary
model, we combine the two constraints on the inflaton potential to
impose an upper limit on $m$, which is severely fine-tuned in the
case of a fixed lattice. We thus conclude that lattice refinement is
necessary to achieve a natural inflationary model.


\end{abstract}

\maketitle

\section{Introduction}
Loop Quantum Gravity (LQG) provides a method of quantising gravity in
a background independent, non-perturbative way.  Quantum gravity
is essential when curvature becomes large, as for example in the
early stages of the evolution of the universe. Applying LQG in a
cosmological context leads to Loop Quantum Cosmology
(LQC)~\cite{Bojowald:2006da}, which is a symmetry reduction of the
infinite dimensional phase space of the full theory, allowing us to
study certain aspects of the theory analytically. In particular, it
has been shown that classical big bang~\cite{Bojowald:2001xe} and
black hole singularities~\cite{Ashtekar:2000eq} are removed in LQC, in
a well defined manner. The discreteness of spatial geometry, a
key element of the full theory, leads to successes in LQC which do not
hold in the Wheeler-De Witt (WDW) quantum cosmology.

In LQC, the quantum evolution is governed by a second order
difference equation, rather than the second order differential
equation of the WDW theory.  This is at the heart of LQC's success at
removing singularities~\cite{Bojowald:2001xe,Ashtekar:2000eq}.  Of
equal importance is the continuum limit of these equations, which can
lead to a behaviour which is qualitatively different from the
classical cosmological evolution~\cite{Ashtekar:2006uz}.  As the
universe becomes large and enters the semi-classical regime, the WDW
differential equation becomes a very good approximation to the
difference equation of LQC.

By deriving the effective equations of motion in the continuum limit,
a semi-classical approach can be used to investigate possible
phenomenological signatures of the theory. In particular, it was hoped
that LQC would help overcome the extreme fine tuning necessary to
achieve successful inflation in general
relativity~\cite{Gibbons:2006pa}.  However, it has been
shown~\cite{Germani:2007rt} that semi-classical corrections are
insufficient to alleviate this difficulty. There remains the
possibility that effects due to the underlying discrete nature of LQC
may come to inflation's aid. Here, we will explore this possibility
and show that the discrete nature of the early universe naturally
introduces constraints on a subsequent inflationary era.

It has been suggested, by heuristic considerations, that for LQC to be
considered physical the scale at which the discrete structure of
space-time becomes significant must shrink as the universe
expands~\cite{Vandersloot PhD}. This lattice refinement has been put
on a more rigorous footing in Ref.~\cite{Bojowald:2007ra}.  In
particular, Ref.~\cite{Bojowald:2007ra} has analysed implications of
lattice refinement for the semi-classical behaviour of the dynamical
difference equations.  Studying the effect of different choices for
the lattice spacings on the solutions of the Hamiltonian constraint,
the authors of Ref.~\cite{Bojowald:2007ra} deduced the stability
properties of the various refinement models. Combining the stability
analysis with a confirmation of the appropriate classical limit, the
suitable refinement models were constrained~\cite{Bojowald:2007ra}.

In what follows, in addition to discussing the effects of LQC on an
inflationary era, we will show that lattice refinement is
necessary to support a massive scalar field (in our context the
inflaton), and we will generalise to matter components with an
arbitrary scale factor dependence.  In particular, we will argue that
lattice refinement renders a successful inflationary era more natural.

In Section 2, we briefly discuss elements of LQC which we will later
use. We are interested in isotropic models described in terms of one
dynamical parameter. In Section 3, we quantise the theory and derive
the Hamiltonian constraint. We then take its continuum limit, assuming
that the regulating length (which is proportional to the length of the
holonomy) is constant.  We also discuss quantisation ambiguities.  In
Section 4, we investigate solutions of the full Hamiltonian constraint
for large scales, which is indeed the WDW equation for a
massive scalar field. In Section 4, we discuss lattice refinement and
we deduce the constraints imposed on the inflationary potential. In
Section 5, we briefly discuss non-flat geometries. We round up our
conclusions in Section 6.

\section{Elements of Loop Quantum Cosmology}\label{sec:LQC}
We deal with isotropic models described in terms of the scalar field
$a(t)$, thus there is only one kinematical degree of freedom. In
connection variables, it is parametrised by the triad component
$\tilde p$ conjugate to the connection component $\tilde c$. Their
relation to the metric variable $a$ is~\cite{Ashtekar:2003hd}
\be 
|\tilde p|=a^2~~~~~ \mbox{and}~~~~ \tilde
c=k+\gamma\dot a~, 
\ee
where over-dot denotes derivatives w.r.t. cosmological time, $\gamma$
is the Barbero-Immirzi parameter, representing a quantum ambiguity
parameter of the theory, and $k$ depends on the intrinsic spatial
curvature ($k=0,\pm 1$).  We will concentrate on flat universes, thus
$k=0$.  The triad has an orientation, determined by the sign of
$\tilde p$, and since the scale factor has no orientation, we used an
absolute value in the above equation.  The triad $\tilde p$
incorporates information about the three-geometry through the scale
factor $a$, while the connection $\tilde c$ has information about the
curvature, which in our case (flat and isotropic geometry) is
contained in the extrinsic curvature proportional to $\dot a$.  The
canonical variables $\tilde c, \tilde p$ are related through
\be
\{\tilde c, \tilde p\}=\frac{\kappa\gamma}{3} V_0~,
\ee
where $\kappa\equiv 8\pi G$ and $V_0$ is the volume of the elementary
cell ${\cal V}$ adapted to the fiducial triad. Defining
\be
p=V_0^{2/3}\tilde p ~~~~ \mbox{and}~~~~ c=V_0^{1/3}\tilde c~,
\ee
with the triad component $p$ determining the physical volume of the
fiducial cell, and the connection component $c$ determining the 
rate of change of the physical edge length of the fiducial cell, one
obtains
\be
\{c, p\}=\frac{\kappa\gamma}{3}~,
\ee
independent of the volume of the fiducial cell.

To proceed in the quantum theory, one follows the procedure used in
the full LQG theory.  Thus, in LQC one takes $e^{i\mu_0 c/2}$, with
$\mu_0$ an arbitrary real number and $p$ as the elementary classical
variables, which have well-defined operator
analogues~\cite{Ashtekar:2003hd}.

Using the Dirac bra-ket notation and setting $e^{i\mu_0 c/2}=\langle
c|\mu\rangle$, the eigenstates of $\hat p$ are the basis vectors
$|\mu\rangle$:
\be
\hat p|\mu\rangle=\frac{\kappa\gamma\hbar|\mu|}{6}|\mu\rangle~.
\ee
Using the volume operator $\hat{V}=|\hat{p}|^{3/2}$, representing the
volume of the elementary cell ${\cal V}$, with eigenvalues $V_\mu=
(\kappa \gamma \hbar |\mu| / 6)^{3/2}$, one
gets~\cite{Ashtekar:2006uz}
\be 
\label{eq:vol}
\hat{V}|\mu\rangle = \left(\frac{\kappa \gamma \hbar |\mu|}{
 6}\right)^{3/2} |\mu\rangle~.  
\ee 
Notice that since we will be concerned with the large scale behaviour
of the LQC equations, we will typically  neglect the sign ambiguity that
arises because the triads can have two different orientations. We
thus avoid numerous factors of ${\rm sgn}(\mu)$ appearing in
subsequent equations.

To define the inverse volume operator, one has to trace over SU(2)
valued holonomies. Since there is a freedom in choosing the
irreducible representation to perform the trace, an ambiguity, labelled
by $J$, arises. More precisely, the half-integer $J$ stands for the
spin of the representation. At this point, it is important to note
that the same ambiguity arises in the gravitational part of the
Hamiltonian constraint, since one has again to trace over SU(2) valued
holonomies. Usually one quantises the gravitational part of the
Hamiltonian constraint using the fundamental $J=1/2$ representation,
and the ambiguity is only investigated for the matter part. This issue
is further discussed in Section \ref{subsec:amb}.

Let us use the $J=1/2$ irreducible representation of SU(2). The
inverse volume operator is diagonal in the $|\mu\rangle$ basis and is
given by~\cite{Ashtekar:2003hd}
\be\label{eq:inv_vol} 
\widehat{V^{-1}}| \mu\rangle= \left| \frac{6}{\kappa \gamma \hbar \mu_0}
\left( V_{\mu+\mu_0}^{1/3} -V_{\mu - \mu_0}^{1/3} \right) \right|^3
|\mu\rangle~,
\ee
where $\mu_0$ is proportional to the length of the holonomy. At this
point, we would like to stress that the regulating length $\mu_0$ is
the crucial parameter in the quantum corrections. In
Eq.~(\ref{eq:inv_vol}) the eigenvalues are bounded and approach zero
near the classical singularity, in contrast to the
classical case where the eigenvalues diverge at the singularity
$\mu=0$. The
eigenvalues reach their maximum at a characteristic scale equal to
$\mu_0$, while at larger $\mu$ they approach the classical values and
below $\mu_0$ they are suppressed compared to their classical
analogues~\cite{Vandersloot PhD}.  The inverse volume is cutoff for
small volumes and, as it has been shown~\cite{Bojowald:2002ny}, the
generalisation to higher $J$ will push the cutoff region to larger
volumes.

\section{Hamiltonian constraint}\label{subsec:Ham}

The dynamics are determined completely by the Hamiltonian operator, the
gravitational part of which, in the fundamental representation,
is~\cite{Vandersloot PhD},
\be\label{eq:ham}
\hat{\mathcal C}_{\rm g}=\frac{2i}{\kappa^2\hbar\gamma^3\mu_0^3}{\rm
tr} \sum_{ijk} \epsilon^{\rm ijk} \left( \hat{h}_i\hat{h}_
j\hat{h}^{-1}_i
\hat{h}^{-1}_j\hat{h}_k\left[\hat{h}^{-1}_k,\hat{V}\right]\right)
(\hat{p})~,
\ee
where 
\be
\hat{h}_i=\cos\biggr(\frac{\mu_0 c}{2}\biggl)-
i\sigma_i\sin\biggr(\frac{\mu_0 c}{2}\biggl)~,
\ee
denote the holonomies, with $\sigma_{\rm i}$ the Pauli spin matrices;
the trace is taken over the SU(2) indices. We quantise this by setting
\be
\hat{p}|\mu\rangle=p|\mu\rangle ~~\mbox{and} ~~\widehat{ e^{\mu_0
c/2}} |\mu\rangle =e^ {\mu_0 \frac{{\rm d}}{{\rm d}\mu } }|\mu
\rangle=|\mu+\mu_0\rangle~.
\ee  
The action of the self-adjoint Hamiltonian constraint operator,
$\hat{\mathcal H}_{\rm g}=(\hat{\mathcal C}_{\rm g} + \hat{\mathcal
C}_{\rm g}^{\dagger} )/2$, on the basis states, $|\mu\rangle$, is
\be
\hat{\mathcal H}_{\rm g} | \mu \rangle 
=
\frac{3}{4\kappa^2 \gamma^3 \hbar \mu_0^3}
\Bigl\{ \bigl[ R(\mu)+R(\mu+4\mu_0)\bigr]| \mu+4\mu_0 \rangle
-4R(\mu ) | \mu\rangle +\bigl[ R(\mu)+R(\mu-4\mu_0)\bigr]| 
\mu-4\mu_0 \rangle \Bigr\}~,
\ee
where 
\be
R(\mu )= \left( \kappa \gamma \hbar/6\right)^{3/2} 
\Big| | \mu+\mu_0 |^{3/2} -|\mu-\mu_0|^{3/2} \Big|~.
\ee
We impose the constraint equation on the physical wave-functions
$|\Psi\rangle$, which are those annihilated by the constraint operator,
i.e. they have to fulfil the Wheeler-De Witt equation, since the
classical expression must vanish. Such states can be expanded using
the basis states as $|\Psi\rangle = \sum_\mu
\Psi_\mu(\phi)|\mu\rangle$, with summation over values of $\mu$ and
where the dependence of the coefficients on $\phi$ represents the
matter degrees of freedom. Since the states $|\mu\rangle$ are
eigenstates of the triad operator, the coefficients $\Psi_\mu(\phi)$
represent the state in the triad representation.  Thus, quantising the
Friedmann equation along the lines of the constraint in the full LQG
theory, one gets the following difference
equation~\cite{Vandersloot:2005kh}
\beq\label{qee}
&&~~\Biggl[\Big| V_{\mu+5\mu_0}-V_{\mu+3\mu_0}\Big|+\Big|V_{\mu+\mu_0}
 - V_{\mu-\mu_0}\Big|\Biggr] \Psi_{\mu+4\mu_0}(\phi) -
 4\Big|V_{\mu+\mu_0}V_{\mu-\mu_0}\Big| \Psi_\mu(\phi) \nonumber \\
 &&+\Biggl[\Big|V_{\mu-3\mu_0}-
 V_{\mu-5\mu_0}\Big|+\Big|V_{\mu+\mu_0}- V_{\mu-\mu_0}\Big|\Biggr]
 \Psi_{\mu-4\mu_0}(\phi) =- \frac{4\kappa^2 \gamma^3 \hbar \mu_0^3}{3}
 {\mathcal H}_{\rm \phi}(\mu)\Psi_\mu(\phi)~,
\eeq
where the matter Hamiltonian $\hat{\mathcal H}_{\rm \phi}$ is assumed
to act diagonally on the basis states with eigenvalue ${\mathcal
H}_{\rm \phi}(\mu)$. Equation (\ref{qee}) is indeed the quantum
evolution (in internal time $\mu$) equation. There is no continuous
variable (the scale factor in classical cosmology), but a label $\mu$
with discrete steps.  The wave-function $\Psi_\mu(\phi)$ depending on
internal time $\mu$ and matter fields $\phi$ determines the dependence
of matter fields on the evolution of the universe.

\subsection{Continuum limit}\label{sec:cont_lim}

We consider the continuum (i.e. $\mu \gg \mu_0$) limit of the
Hamiltonian constraint operator acting on the physical states, i.e.
$\hat{\mathcal H}_{\rm g}|\Psi\rangle $. Expanding around small
regulating length, $\mu_0$, we get
\be\label{exp}
V_{\mu + \alpha \mu_0} - V_{\mu + \beta \mu_0} =
\frac{3}{2} \left(\frac{\kappa \gamma \hbar}{6}\right)^{3/2} \mu_0
\mu^{1/2} \Biggl[ \alpha - \beta + \frac{\mu_0}{4\mu} \left(
\alpha^2 - \beta^2\right) 
-\frac{\mu_0^2}{24\mu^2}\left( \alpha^3 - \beta^3 \right) + {\mathcal
O} \left( \frac{\mu_0^3}{\mu^3}\right) +\mbox {~higher orders}\Biggr]~.
\ee
From Eqs.~(\ref{qee}) and (\ref{exp}) we obtain, in the small
$\mu_0$ limit, that
\beq
\label{eq:ham_g1} 
&&\frac{9}{2\kappa^2\gamma^3\hbar} \left(\frac{\kappa\gamma\hbar}{6}
\right)^{3/2} \mu_0^{-2}
\mu^{1/2} \Biggl[ \Psi_{\mu+4\mu_0}(\phi)-2\Psi_\mu(\phi)
+\Psi_{\mu-4\mu_0}(\phi)
+\frac{\mu_0}{\mu} \Bigl\{
\Psi_{\mu+4\mu_0}(\phi)-\Psi_{\mu-4\mu_0}(\phi) \Bigr\}\nonumber \\
&&\ \ \ \ \ \ \ -\frac{\mu_0^2}{\mu^2} \Bigl\{
\Psi_{\mu+4\mu_0}(\phi)+\Psi_{\mu-4\mu_0}(\phi) \Bigr\}
-\frac{\mu_0^2}{24\mu^2}\Bigl\{ \Psi_{\mu+4\mu_0}
(\phi)-2\Psi_\mu(\phi)+\Psi_{\mu-4\mu_0}(\phi)\Bigr\} + {\mathcal
O}\left(\frac{\mu_0^3}{\mu^3}\right) \Biggr] \nonumber \\
&=&-{\mathcal H}_{\phi}(\mu)\Psi_\mu(\phi)~.  
\eeq
The above second order difference equation distinguishes the
components of the wave-functions in different lattices of spacing
$4\mu_0$.  Assuming that $\Psi$ does not vary much on scales of the
order of $4\mu_0$ (known as {\sl
pre-classicality}~\cite{Bojowald:2002gz}), one can smoothly
interpolate between the points on the discrete function
$\Psi_\mu(\phi)$ and approximate them by the continuous function
$\Psi(\mu,\phi)$. Under this assumption, the difference equation is
be very well approximated by a differential equation for a continuous
wave-function.  As we will discuss later, it is this assumption that
can break down at large scales, leading to deviations from the
classical behaviour.

Let us consider the parameter $\mu_0$ entering in the regularisation
of the Hamiltonian constraint to be a constant real number (the case
of lattice refinement will be addressed in Section
\ref{subsec:lattice_ref}). In other words, let us assume a fixed
spatial lattice throughout the whole evolution of the
universe. Expanding the wave-function $\Psi(\mu\pm 4\mu_0,\phi)$ in
Taylor series we obtain:
\be
\label{taylor} \Psi(\mu\pm 4\mu_0,\phi)=\Psi(\mu,\phi)
\pm\frac{\partial \Psi(\mu,\phi)}{\partial\mu}(4\mu_0)
+\frac{1}{2}\frac{\partial^2\Psi(\mu,\phi)}{\partial\mu^2}(16\mu_0^2)
+{\mathcal O}\left(\mu_0^3
\frac{\partial^3\Psi(\mu,\phi)}{\partial\mu^3}\right) +{\mathcal
O}\left(\mu_0^4
\frac{\partial^4\Psi(\mu,\phi)}{\partial\mu^4}\right)+\cdots~.  
\ee
Applying Eq.~(\ref{taylor}) in the difference equation
Eq.~(\ref{eq:ham_g1}) we get 
\be
\label{eq:ham_g2}
\frac{36}{\kappa^2\gamma^3\hbar} \left(\frac{\kappa\gamma\hbar}{6}
\right)^{3/2} \Biggl[ \sqrt{\mu}\frac{\partial^2 \Psi(\mu,
\phi)}{\partial \mu^2}+\frac{\partial^2 }{ \partial \mu^2}
\Bigl(\sqrt{\mu}\Psi(\mu,\phi)\Bigr)\Biggr] +{\mathcal
O}\left(\mu_0\right) + {\mathcal O}\left(\mu_0^2\right)+\cdots
=-{\mathcal H}_{\phi}(\mu)\sqrt\mu \Psi(\mu,\phi)~,
\ee 
where $\cdots$ denote higher order corrections in $\mu_0$.
Classically, the matter part of the Hamiltonian constraint for a
massive scalar field is just, 
\be 
\label{eq:ham_m1} 
{\mathcal H}_\phi= \kappa\left[\frac{P_\phi^2}{2a^3} +a^3 V(\phi)
\right]~, 
\ee 
with momentum $P_\phi$ and potential $V(\phi)$.  To quantise the above
Hamiltonian constraint we use
\be 
\hat{P}_\phi \Psi(p,\phi) = -i\hbar
\frac{\partial \Psi(p,\phi)}{\partial\phi}\ \ \ \ \mbox{and}\ \ \ \
\hat{\phi}\Psi(p,\phi) = \phi \Psi(p,\phi)~, 
\ee 
while the expressions for the volume and inverse volume operators,
$\hat{V}$ and $\widehat{V^{-1}}$, are given by Eqs.~(\ref{eq:vol}) and
(\ref{eq:inv_vol}), respectively.  Taking then the continuum limit,
one obtains
\be
\label{conti} \hat{\mathcal H}_{\phi} \Psi(\mu,\phi)=
-3\left( \frac{6\hbar}{\kappa \gamma^3}\right)^{1/2} \mu^{-3/2}
\frac{\partial^2\Psi(\mu,\phi)}{\partial \phi^2} + \left( \frac{\kappa
\gamma \hbar}{6} \right)^{3/2} \mu^{3/2} V(\phi) \Psi(\mu,\phi)
+{\mathcal O}\left(\mu_0\right)+\cdots~.
\ee 
From
Eqs.~(\ref{eq:ham_g2}) and (\ref{conti}), the full Hamiltonian
constraint, $\left( \hat{\mathcal H}_{\rm g} + \hat{\mathcal H}_{\phi}
\right) |\Psi \rangle =0$, reads 
\be 
\label{eq:WdW0}
\sqrt{\mu}\frac{\partial^2\Psi(\mu,\phi)}{\partial\mu^2}
+\frac{\partial^2}{\partial \mu^2} \left( \sqrt{\mu}
\Psi(\mu,\phi)\right) - 3\mu^{-3/2}\frac{ \partial^2
\Psi(\mu,\phi)}{\partial\phi^2} +\frac{\kappa^2 \gamma^3 \hbar}{36}
V(\phi) \mu^{3/2} \Psi(\mu,\phi) +{\mathcal
O}\left(\mu_0\right)+\cdots=0~, 
\ee 
or, equivalently, in terms of $p$ with 
\be
\label{p}
p=\kappa\gamma\hbar\mu/6~, 
\ee 
one obtains 
\be
\label{eq:WdW1} 
\sqrt{p}\frac{\partial^2 \Psi(p,\phi)}{\partial p^2}
+\frac{\partial^2}{\partial p^2} \left(\sqrt{p}\Psi(p,\phi)\right) - 3
p^{-3/2} \frac{\partial^2 \Psi(p,\phi)}{\partial \phi^2}
+\frac{6}{\kappa \hbar^2}p^{3/2} V(\phi) \Psi(p,\phi) +{\mathcal
O}\left(\mu_0\right)+\cdots = 0~, 
\ee 
which is just a particular factor ordering of the Wheeler-DeWitt
equation for a massive scalar field.

\subsection{Ambiguities}\label{subsec:amb}

Thus far we have neglected any quantisation ambiguities, which can
have significant consequences for the predictions of
LQC~\cite{Germani:2007rt},~\cite{Bojowald:2002ny}. As it was discussed
in Section 2, there is a fundamental ambiguity arising from our choice
of representation for the holonomies.  It is possible to account for
this ambiguity in the eigenvalues of the inverse volume
operator~\cite{Bojowald:2002ny}, however constructing the
gravitational part of the Hamiltonian constraint in non-fundamental
($J\neq 1/2$) representations, results in a higher order difference
equation~\cite{Vandersloot PhD}. For $J=1$ the difference equation has
fundamental step size equal to $2\mu_0$, and it is of order 
$8J=8$~\cite{Vandersloot:2005kh}.  Being of higher order, spurious
solutions may arise~\cite{Vandersloot:2005kh} suggesting that representations
of higher order than the fundamental representation may not
lead to the correct quantum theory.  Whilst it may appear more natural
to choose the same representation for both the matter and
gravitational parts of the Hamiltonian constraint, there is no {\it a
priori} reason this should be the case.  Here we set $J=1/2$ in the
gravitational part of the constraint since it is the most tractable,
whilst leaving it unspecified in the matter part, to remain as general
as possible.

The $J$ ambiguity arises only in the inverse volume eigenvalues,
which are given in general by~\cite{Vandersloot PhD}
\be\label{ambJL}
\widehat{V^{-1}}|\mu\rangle = \left| \frac{9}{\kappa \gamma \hbar
\mu_0 L J\left(J+1 \right)\left( 2J+1\right)} \sum_{m=-J}^{J}
mV^{2L/3}_{\mu+2m\mu_0} \right|^{\frac{3}{2(1-L)}}|\mu\rangle~,
\ee
where $0<L<1$ is a second quantisation ambiguity coming from
the classical Poisson bracket~\cite{Vandersloot PhD}
\be
\{ c,|p|^L \}=\frac{1}{3}\gamma \kappa L {\rm sgn}(p)|p|^{L-1}~.
\ee
Taking the $\mu>>\mu_0$ limit, Eq.~(\ref{ambJL}) implies
\be \label{eq:vol_inv} 
\widehat{V^{-1}} |\mu\rangle = \left[ 1 +
\frac{(L-2)(3J^2+3J-1)\mu_0^2}{5\mu^2}+ {\mathcal
O}\left(\frac{\mu_0^4}{\mu^4}\right)+\cdots \right]|p|^{-3/2}|\mu\rangle~.
 \ee
There are also ambiguities arising from the classical matter
Hamiltonian we choose to
quantise~\cite{Germani:2007rt},~\cite{Bojowald:2004xq}.
Classically,
\be \label{eq:ham_m2}
 {\mathcal H}_\phi=
 \kappa\left[\frac{P_\phi^2}{2a^{3(n+1)}}a^{3n}
 +\frac{1}{a^{3m}}a^{3(m+1)} V(\phi) \right]~,
\ee
is identical to Eq.~(\ref{eq:ham_m1}), however they will differ in the
quantum regime. 
Equation (\ref{eq:vol_inv}) implies 
\be
\label{amb2}
\left(\hat{V}\widehat{V^{-1}}\right)^n |\mu\rangle = \left[ 1 +
\frac{n(L-2)(3J^2+3J-1)\mu_0^2}{ 5\mu^2} + {\mathcal O}\left(
\frac{\mu_0^4}{\mu^4}\right) +\cdots \right] |\mu\rangle~.  
\ee 
Equations (\ref{eq:vol_inv}) and (\ref{amb2}) imply that all quantum
ambiguities are of order ${\mathcal O}\left( \mu_0^2 / \mu^2 \right)$,
thus they do not affect the results of Section \ref{sec:cont_lim}.

\subsection{Wheeler-De Witt Solutions}\label{subsec:WdW}

From Eq.~(\ref{eq:WdW1}) we see that for large scales the equation we
want to solve is the standard Wheeler-De Witt equation,
\be \label{eq:WdW2}
\sqrt{p}\frac{\partial^2 \Psi(p,\phi)}{\partial p^2}
+\frac{\partial^2}{\partial p^2} \left( \sqrt{p} \Psi(p,\phi)\right)
 + \alpha p^{3/2} V(\phi)\Psi(p,\phi)=0~,
\ee
where 
\be \label{alpha}
\alpha=\frac{6}{\kappa \hbar^2}~.
\ee
At this point we approximate the dynamics of $\phi$ to be such that
\be
V(\phi)=p^{\delta -3/2}V_\phi~,
\ee 
with $V_\phi$ a constant.  For an inflaton field during slow-roll
inflation, $V(\phi)\approx $const., so we expect $\delta \approx 3/2$
in this era (which should be during the continuum epoch to be compatible
with observations). By separation of variables
$\Psi(p,\phi)=\Upsilon(p)\Phi(\phi)$, we get
\be\label{wdwT}
\sqrt{p}\frac{{\rm d}^2 \Upsilon(p)}{{\rm d}p^2}
+\frac{{\rm d}^2 }{{\rm d}p^2} \left( \sqrt{p} \ \Upsilon(p) \right)
 + \alpha p^\delta V_\phi \Upsilon(p) \approx 0 ~.
\ee 
Equation (\ref{wdwT}) has solutions
\be
\Upsilon(p)=C_1 p^{1/4} \mbox{\large
\it{J}}_{\frac{\sqrt{3}}{2\delta+3}} \left(
\frac{2\sqrt{2\alpha}\sqrt{V_\phi}}{2\delta+3}
p^{(2\delta+3)/4}\right)
 + C_2 p^{1/4} \mbox{\large
\it{Y}}_{\frac{\sqrt{3}}{2\delta+3}} \left( \frac{ 2
\sqrt{2\alpha}\sqrt{V_\phi} }{2\delta+3}p^{(2\delta+3)/4}\right)~,
\ee
where $J$ and $Y$ are Bessel functions of the first and second kind,
respectively, and $C_1$ and $C_2$ are integration constants.  It is
clear that for $\delta >1/2$, the period of the solution 
decreases with increasing $\mu$. Without lattice refinement (to be
discussed in the next section) the discrete nature of the underlying
lattice would eventually be unable to support the oscillations and the
assumption of pre-classicality will break down, implying that the
discrete nature of space time becomes significant on very large
scales.

We can see this rigorously by using the Bessel function expansions,
\beq
 \mbox{ \large \it{J}}_\beta (x) &\rightarrow& 
\sqrt{\frac{2}{\pi x}} \cos \left( x -
 \frac{\beta \pi}{2} - \frac{\pi}{4} \right)~, \nonumber \\
 \mbox{ \large \it{Y}}_\beta
 (x) &\rightarrow& \sqrt{\frac{2}{\pi x}} \sin \left( x - \frac{\beta
 \pi}{2} - \frac{\pi}{4} \right)~, \nonumber
\eeq
for large $x$, which give
\be 
\Upsilon(p) \approx \sqrt{\frac{2\delta+3}{\pi \sqrt{2\alpha}\sqrt{V_\phi}}}
 p^{-(2\delta+1)/8} \Biggl[ C_1 \cos\left( x -
 \frac{\sqrt{3}\pi}{2(2\delta+3)} - \frac{\pi}{4} \right)+C_2
 \sin\left( x - \frac{\sqrt{3}\pi}{2(2\delta+3)} - \frac{\pi}{4} \right)
 \Biggr]~,
\ee
where $x= 2\sqrt{2\alpha}\sqrt{V_\phi}(2\delta+3)^{-1}
p^{(2\delta+3)/4}$.  The zeros of $\Upsilon(p)$ are at
\be
\tan \left( x - \frac{\pi}{2(\delta+2)}-\frac{\pi}{4} 
\right) = -\frac{C_1}{C_2}~,
\ee
which gives the scale of the $n^{th}$-zero to be, 
\be 
p_n=\left[\frac{(2\delta+3)\pi}{2\sqrt{2\alpha}
\sqrt{V_\phi}}\right]^{\frac{4}{2\delta+3}} \left[
C + n\right] ^{\frac{4}{2\delta+3}}~, 
\ee 
with
\be
C=\tan^{-1} \left(-\frac{C_1}{C_2}\right) \frac{1}{\pi} +
\frac{\sqrt{3}}{2(2\delta+3)} \pm\frac{1}{2} -\frac{1}{4}~,
\ee
a constant. Note that we take the $n^{th}$ zero of $\tan (x)$ to 
be at $x=(2n\pm 1)\pi/2$. 

We are interested in the separation between two successive zeros in
the limit of large $n$ (large $p$), which is given by,
\be\label{eq:sep1} 
\Delta p \equiv \lim_{n\rightarrow \infty}\left(p_{n+1}-p_n\right) =
\frac{2\pi}{\sqrt{2\alpha}\sqrt{V_\phi}} p^{(1-2\delta)/4}+{\mathcal
O}\left(p^{-(2\delta+1)/2}\right)~.  
\ee
For $\mu_0$ constant, it is clear that for $\alpha \neq 0$ 
($\alpha =0$ corresponds to the case where the wave-function
is linear and hence has an infinite period), there will be a scale at
which $\Delta p$ is of the order of $2\kappa\gamma\hbar\mu_0/3$ (i.e.
when $\Delta \mu$ is of the order of $4\mu_0$), at which point the
assumption that the wave-function is smooth on scales of the order of
$4\mu_0$  breaks down and the semi-classical description is no
longer valid. If we want the end of inflation to be describable using
classical general relativity then it must end before this scale is
reached. During slow-roll inflation the scalar field potential is
approximately constant so we can set $\delta \approx 3/2$.

For $\Delta p>2\kappa\gamma \hbar\mu_0/3$,
Eqs.~(\ref{p}),~(\ref{alpha}) and (\ref{eq:sep1}) lead to the following
constraint for the scale, $a_f$, at the end of inflation:
\be\label{afinal}
 a_{\rm f} < \frac{3\pi}{2\gamma\mu_0 \sqrt{ 3\kappa V_\phi}}~,
\ee
where we have also used that $a=|p|^{1/2}=(\kappa\gamma\hbar|\mu|/6)^{1/2}$.
We can then calculate the maximum number of e-foldings, $N_{\rm cl}=
\ln(a_{\rm f}/a_{\rm i})$, where $a_i$ is the initial scale, which can
occur within the classical era, given by
\be\label{ainitial}
a_{\rm i} \gg \left(2\kappa \gamma \hbar\mu_0/3\right)^{1/2}~.
\ee
Thus, from Eqs.~(\ref{afinal}) and (\ref{ainitial}) the number of
e-foldings $N_{\rm cl}$ must satisfy the constraint
\be\label{efolds}
N_{\rm cl} \ll \ln \left( \frac{3\pi}{2 \kappa \sqrt{ 2 V_\phi \hbar \gamma^3
\mu_0^3}} \right)~.
\ee
Setting~\cite{Ashtekar:2006uz} $\mu_0=3\sqrt{3}/2$ and $\gamma\approx
0.24$, the above constraint, Eq.~(\ref{efolds}), on the number of
e-foldings in units of $\hbar=1$ reads
\be
 N_{\rm cl} \ll \ln \left(\frac{0.27}{\sqrt{V_\phi}l_{\rm Pl}^2}\right)~,
\ee
with $l_{\rm Pl}=\sqrt{8\pi G}$.  Thus, to have $N_{\rm cl}$
e-foldings of inflation with negligible quantum gravity effects,
the potential must satisfy
\be\label{constrVfixed}
V_\phi \ll 0.07e^{-2N_{\rm cl}}l_{\rm pl}^{-4}~.
\ee
If we, conservatively, say that only half of inflation takes place 
in the classical era, the above constraint leads to
\be
V_\phi \ll 10^{-28}l_{\rm pl}^{-4}~.
\ee
Clearly, the above constraint becomes even tighter if we increase the
number of e-foldings during the classical era.  Considering lattice
refinement we will see that this strong constraint on the value of the
inflaton potential (during slow roll) becomes considerably less
fine-tuned.

\section{Lattice refinement}\label{subsec:lattice_ref}

In the derivation of the Hamiltonian constraint we have assumed that
the parameter $\mu_0$ is a constant real number. This parameter is
related to the edge length used in holonomies. Since holonomies refer
to embedded edges and loops, a constant $\mu_0$ corresponds to a fixed
spatial lattice, so that only edges of a given coordinate length are
used, and consequently the loop size remains the same as the total
volume increases. In what follows we relax the assumption of a
constant $\mu_0$ and we investigate the consequences, and thus the
necessity, of a parameter $\tilde \mu$ being dependent on the phase
space variable $p$, or equivalently, on $\mu$.
                 
Allowing the length scale of the holonomies to vary dynamically, the
form of the difference equation, Eq.~(\ref{qee}), changes.  Since the
parameter $\mu_0$ determines the step-size of the difference equation,
assuming the lattice size is growing, the step-size of the difference
equation is not constant in the original triad variables.  The exact
form of the difference equation depends on the lattice refinement
used.  Here we will take the particular case of 
\be 
\mu_0 \rightarrow \tilde{\mu}\left(\mu\right)=\mu_0\mu^{-1/2}~.  
\ee
This type of lattice refinement is suggested by certain intuitive
heuristic approaches such as noting that the {\it minimum area} used
to regulate the holonomies should be a {\it physical}
area~\cite{Ashtekar:2006uz2}, or that the discrete step size of the
difference equation should always be of the order of the Planck
volume~\cite{Vandersloot PhD}. This choice also results in a
significant simplification of the difference equation, compared to
more general lattice refinement schemes.

The basic operators are given by replacing $\mu_0$ with $\tilde{\mu}$.
Upon quantisation we have~\cite{Ashtekar:2006uz2}
\be
\widehat{ e^{i\tilde{\mu}c/2}}
|\mu \rangle=e^{-i\tilde{\mu}\frac{\rm d}{{\rm d}\mu} }|\mu\rangle~,
\ee
which is no longer a simple shift operator since $\tilde{\mu}$ is a 
function of $\mu$. If we change our basis to 
\be\label{eq:basis}
 \nu = \mu_0 \int \frac{{\rm d}\mu}{\tilde{\mu}}=\frac{2}{3} \mu^{3/2}~,
\ee
we find
\be
 e^{-i\tilde{\mu}\frac{\rm d}{{\rm d} \mu}}|\nu\rangle 
= e^{-i\mu_0 \frac{\rm d}{
{\rm d}\nu}}|\nu\rangle = |\nu+\mu_0\rangle~.
\ee
Thus, the basis $|\nu\rangle$ is a much more natural choice than
$|\mu\rangle$. The action of the volume operator on these basis states
is
\be
 \hat{V}|\nu\rangle 
= \frac{3\nu}{2}\left(\frac{\kappa\gamma\hbar}{6}\right)^{3/2}
|\nu\rangle~,
\ee
and the self-adjoint Hamiltonian constraint operator acts
as~\cite{Vandersloot PhD},
\be
 \hat{\mathcal H}_{\rm g} |\nu\rangle = \frac{9|\nu|}{16\mu_0^3} \left(
\frac{\hbar}{6\kappa\gamma^3}\right)^{1/2}
 \Biggl[ \frac{1}{2}\Bigl\{ U\left( \nu \right) 
+U\left( \nu+4\mu_0 \right) 
\Bigr\} |\nu+ 4\mu_0\rangle -2U\left( \nu \right) |\nu \rangle 
+ \frac{1}{2} \Bigl\{ U\left( \nu \right) + U\left( \nu-4\mu_0\right) 
\Bigr\} |\nu - 4 \mu_0 \rangle \Biggr]~,
\ee
where 
\be
U\left(\nu\right) = |\nu + \mu_0 | - | \nu - \mu_0|~.
\ee
Notice that since $\mu$ is dimensionless, so is $\nu$.

Following the same approach as previously, we expand 
$|\Psi\rangle=\sum_\nu \Psi_\nu(\phi) |\nu\rangle$
and the Hamiltonian constraint becomes,
\beq\label{eq:const_refine}
&&  \frac{1}{2} \Big|\nu + 4\mu_0 \Big|
\Bigl[ U\left(\nu + 4\mu_0\right) +U\left( \nu\right) \Bigr]  
\Psi_{\nu+4\mu_0}\left(
\phi \right) 
+ 2|\nu| U\left(\nu\right) \Psi_\nu \left(\nu\right)  
+\frac{1}{2} \Big|\nu-4\mu_0\Big| \Bigl[ U\left(\nu-4\mu_0\right) +
 U\left( \nu\right)
\Bigr] \Psi_{\nu-4\mu_0}\left( \phi \right) \nonumber \\
&& = 
- \frac{16\mu_0^3}{9}\left(\frac{6\kappa\gamma^3}{\hbar}\right)^{1/2}
 {\mathcal H}_{\phi}\left(\nu\right)
 \Psi_{\nu} \left( \phi\right)~.
\eeq
Now we take the continuum ($\nu \gg \mu_0$) limit of
Eq.~(\ref{eq:const_refine}) to get
\be\label{continuumlim}
 \frac{\partial^2 }{\partial \nu^2} 
\Bigl[\nu\Psi\left(\nu,\phi\right)\Bigr]+
 \frac{16}{9}\left(\frac{6\kappa\gamma^3}{\hbar}\right)^{1/2}
 {\mathcal H}_{\phi}\left(\nu\right)
\Psi(\nu,\phi) \nonumber +{\mathcal O}\left( \mu_0\right) + \dots=0~.
\ee
In terms of $\mu$ Eq.~(\ref{continuumlim}) reads
\be
  \mu^{-1/2}\frac{\partial}{\partial \mu} \left[ \mu^{-1/2} 
\frac{\partial}{\partial \mu } \left( \mu^{3/2} \Psi\left(\mu,\phi\right)
 \right) \right]
+\frac{8}{3}\left(\frac{6\kappa\gamma^3}{\hbar}\right)^{1/2}
 {\mathcal H}_{\phi}\left(\mu\right) \Psi\left(\mu,\phi\right) 
+{\mathcal O}\left( \mu_0\right) + \dots=0~,
\ee
which is just a different factor ordering of the Wheeler-DeWitt
equation, Eq.~(\ref{eq:WdW0}). Substituting in for ${\mathcal H}_{\phi}$
from Eq.~(\ref{conti}), and for $p$ from Eq.~(\ref{p}), we get
\be
p^{-1/2}\frac{\partial}{\partial p} \left[ p^{-1/2} 
\frac{\partial}{\partial p } \left( p^{3/2} 
\Psi\left(p,\phi\right) \right) \right]
-48p^{-3/2}\frac{\partial^2\Psi(p,\phi)}{\partial\phi^2} 
+ \frac{96}{\kappa \hbar^2} V(\phi) p^{3/2}\Psi(p,\phi) +
 {\mathcal O}\left(\mu_0\right) + \dots =0~.
\ee
Thus, in the large scale limit the equation we have to solve reads
\be\label{eq:ham_con_refine}
 p^{-1/2}\frac{\partial}{\partial p} \left[ p^{-1/2} 
\frac{\partial}{\partial p } \left( p^{3/2} 
\Psi\left(p,\phi\right) \right) \right]
+ \beta V(\phi) p^{3/2}\Psi(p,\phi) =0~,
\ee
where 
\be
\label{beta}
\beta=\frac{96}{\kappa \hbar^2}~.
\ee 
This equation is analogous to Eq.~(\ref{eq:WdW1}) and thus we are
following the same procedure as in Section \ref{subsec:WdW}. We first
separate the wave-function $\Psi(p,\phi)$ into $\Psi(p,\phi)=
\Upsilon(p)\Phi(\phi)$ and we then approximate the dynamics of the
inflaton field by setting $V(\phi) =V_\phi p^{\delta-3/2}$ to get,
\be\label{eq-p}
 p^{-1/2}\frac{{\rm d}}{{\rm d} p} \left[ p^{-1/2} 
\frac{{\rm d}}{{\rm d} p } \left( p^{3/2} 
\Upsilon\left(p\right) \right) \right]
+ \beta V_\phi p^\delta\Upsilon(p) =0~.
\ee
The solutions of Eq.~(\ref{eq-p}) are
\be\label{bessel-ref}
\Upsilon\left( p\right) 
= C_1p^{-3/4} \mbox{ \large \it{J}}_{\frac{3}{2\delta+3}}
\left(\frac{4\sqrt{\beta V_\phi}}{ 2\delta+3}p^{(2\delta+3)/4}\right)
+ C_2 p^{-3/4} \mbox{\large \it{Y} }_{\frac{3}{2\delta+3}}
\left(\frac{4\sqrt{\beta V_\phi}}{ 2\delta+3}p^{(2\delta+3)/4}\right)~.
\ee
Expanding Eq.~(\ref{bessel-ref}), as we did in Section
\ref{subsec:WdW}, we obtain
\be
 \Upsilon\left( p\right) \approx  p^{-(9+2\delta)/8} \sqrt{\frac{
2\delta+3}{2\sqrt{\beta V_\phi}\pi} }
\Biggl[ C_1 \cos \left( x-\frac{3\pi}{2(2\delta+3)}
-\frac{\pi}{4} \right)
+C_2 \sin \left( x -\frac{3\pi}{2(2\delta+3)}-\frac{\pi}{4} \right)
\Biggr]~,
\ee
where 
\be 
x=4\sqrt{\beta V_\phi}(2\delta+3)^{-1} p^{(2\delta+3)/4}~,
\ee 
with $\beta$ defined in Eq.~(\ref{beta}).
The scale of the $n^{th}$ zero is
\be
 p_n = \left[\frac{(2\delta+3)\pi}
{4\sqrt{\beta V_\phi}}\right]^{\frac{4}{2\delta+3}}
\left[ C+n\right]^{\frac{4}{2\delta+3}}~,
\ee
where
\be
 C=\tan^{-1} \left(\frac{-C_1}{C_2}\right) 
\frac{1}{\pi}+\frac{3}{2(2\delta+3)}
 \pm \frac{1}{2} -\frac{1}{4}~.
\ee
Calculating the separation between two successive zeros, we obtain
\be\label{Deltap}
 \Delta p \equiv \lim_{p\rightarrow \infty}\left(p_{n+1}-p_n\right)
=\frac{\pi}{\sqrt{\beta V_\phi}} p^{(1-2\delta)/4}~.
\ee
For the continuum limit to be valid, the wave-function must vary
slowly on scales of the order of $\mu_c = 4\tilde{\mu}$. Thus, using 
$p=\kappa\gamma\hbar\mu/6$, we arrive at the constraint,
\be\label{constr-p}
\Delta p > 4\mu_0\left(\frac{\kappa\gamma\hbar}{6}\right)^{3/2}p^{-1/2}~.
\ee
Form Eqs.~(\ref{Deltap}) and (\ref{constr-p}) we obtain the following
constraint on $V_\phi$,
\be
 V_\phi < \frac{27\pi^2}{192\mu_0^2\gamma^3 \kappa^2\hbar} 
p^{(3-2\delta)/2}~.
\ee
For slow-roll inflation, $V(\phi)$ must be approximately constant,
thus $\delta\approx 3/2$. As previously, we use $\mu_0=3\sqrt{3}/2$ and
$\gamma\approx 0.24$, 
to find the constraint on the inflationary potential in units of
$\hbar=1$,
\be
 V(\phi) \lsim2.35\times10^{-2} l_{\rm Pl}^{-4}~,
\ee
where $l_{\rm Pl}^2=G\hbar$.  Clearly, this is a milder constraint than
the one obtained for fixed lattices, Eq.~(\ref{constrVfixed}).

\section{Open/closed ($k=\pm 1$) geometries}

In the previous sections we looked at the $k=0$ case only. The reason
for this is that there are several conceptual difficulties associated
with the quantisation procedure in the other cases,
particularly for $k=-1$ ~\cite{Vandersloot:2005_2,Ashtekar:2007}. These
difficulties arise because vectors no longer commute and hence the
four holonomies previously used to regulate the Hamiltonian no longer
form a closed loop.  There are methods of dealing with this (see
Ref.~\cite{Vandersloot PhD} and references therein) that lead to
discrete evolution equations of the same form as
Eqs.~(\ref{qee}) and (\ref{eq:const_refine}), but with an extra term
involving the curvature. 

The correct quantisation should be able to reproduce the
Wheeler-DeWitt equation including a curvature term in a suitable limit,
\be\label{eq:WdW_k}
 \frac{\partial^2 S(\mu,\phi)}{\partial \mu^2} +kS(\mu,\phi) + 
\tilde{\alpha}\mu^\delta S(\mu,\phi)=0~,
\ee
where $S(\mu,\phi)=\sqrt{\mu}\Psi(\mu,\phi)$.
This no longer has general solutions, however during inflation when
$\delta \approx 1$, analytic solutions do exist,
\be
T(\mu) = C_1 {\rm Airy_A}\left(
-\left(k+\alpha\mu\right)\alpha^{-2/3} \right) + C_2 {\rm
Airy_B}\left( -\left(k+\alpha\mu\right)\alpha^{-2/3} \right)~,
\ee
where $T(\mu)$ is the $\mu$ dependent part of $S(\mu,\phi)$
(i.e. $S(\mu,\phi)=T(\mu)\Phi(\phi)$).  Using the Airy functions
expansions,
\beq
\lim_{x\rightarrow \infty} {\rm Airy_A}\left(
-|x|\right)&=&\frac{1}{\sqrt{\pi}x^{1/4}}\sin\left(\frac{2}{3} x^{2/3}
+ \frac{\pi}{4} \right)~, \nonumber \\
 \lim_{x\rightarrow \infty} {\rm
Airy_B}\left(
-|x|\right)&=&\frac{1}{\sqrt{\pi}x^{1/4}}\cos\left(\frac{2}{3} x^{2/3}
+ \frac{\pi}{4} \right)~, \nonumber
\eeq
we obtain  an explicit form for $T(\mu)$, which has zeros at the scales
\be
\mu_n=\alpha^{-1/3} \left(\frac{3\pi}{2} \right)^{2/3}\left( C +
n\right)^{2/3} -\frac{k}{\alpha}~,
\ee
where now 
\be
C=\tan^{-1}\left( -C_2/C_1\right) -3/4~,
\ee
is again constant. Clearly then, the presence of a curvature will not
affect the large scale behaviour of the scale between subsequent zeros
and so we expect the previous section to hold for any quantisation
schemes that have a continuum limit of the form of
Eq.~(\ref{eq:WdW_k}). As we have seen, changing the factor ordering of
the Wheeler-DeWitt equation often means that analytical solutions are
not available, in particular adding a $k\sqrt{p}\ \Upsilon(p)$ term to
Eq.~(\ref{eq:ham_con_refine}) results in an equation that cannot be
solved exactly. However, the success of this factor ordering (which
arises from a non self-adjoint Hamiltonian constraint) suggests that
the results for other cases may be similar.

\section{Discussion/Conclusions}

We have studied phenomenological aspects of LQC.  In particular, we
have investigated the effect of lattice refinement, which as we will
now argue, makes the conditions for successful inflation more natural.

Let us consider that the scalar field $\phi$ plays the r\^ole of the
inflaton. We will then combine the condition imposed earlier on the
potential $V(\phi)$, with the constraint so that the perturbations
induced by $\phi$ are consistent with the CMB measurements on large
angular scales. As a concrete example we will use a simple
inflationary model, namely $V(\phi)=m^2\phi^2/2$.

The contribution of the inflaton field $\phi$ with potential $V(\phi)$
to the fractional over-density in Fourier space, at the epoch of
horizon crossing, $k=aH$, is given by
\begin{equation}
\label{deltaH}
 \delta^2_H\left(k\right) = \frac{1}{75 \pi^2 M_{\rm Pl}^6}
\frac{V^3(\phi)}{ \left[V'(\phi)\right]^2} \Biggl|_{k=aH}~,
\end{equation}
where $V'(\phi)\equiv{\rm d}V/{\rm d}\phi$ and the Planck mass is
$M_{\rm Pl}=(8\pi G)^{-1/2}\simeq 2.4\times 10^{18}{\rm GeV}$.  From the
COBE-DMR measurements, $\delta_H\approx1.91\times10^{-5}$~\cite{cobe},
which implies from Eq.~(\ref{deltaH}) that
\begin{equation}
\label{cobe}
 \frac{\left[V(\phi)\right]^{3/2}}{V'(\phi)}
\approx 5.2\times 10^{-4} M_{\rm Pl}^3~.
\end{equation}
To ensure the continuum approximation is not broken at large scales,
we found in previous sections (Sections III and IV) that the
potential must be $V(\phi)\lsim x$, where $x$ depends on whether we
have static or dynamically varying holonomy length scales. In the case
of fixed lattice and assuming $N_{\rm cl}$ e-foldings of inflation
within the classical era, we found $x^{\rm fixed ~lattice} \sim
7\times 10^{-2} e^{-2N_{\rm cl}} l_{\rm pl}^{-4}$. Whilst, considering
lattice refinement we got $x^{\rm lattice ~refinement}(\phi) \lsim 10^{-2}
l_{\rm Pl}^{-4}$.

We now consider the simple inflationary case $V(\phi)= m^2\phi^2/2$,
favoured by the latest WMAP measurements. In this
case, Eq.~(\ref{cobe}) implies
\begin{equation}\label{eq:const1}
 m\phi^2 \approx 1.5\times 10^{-3}M_{\rm Pl}^{3}~,
\end{equation}
and the constraint on the potential so that the continuum
approximation is valid on large scales reads
\begin{equation}
\label{eq:const2}
 m^2\phi^2 \lsim 2x~.
\end{equation}
Equations (\ref{eq:const1}) and (\ref{eq:const2}) imply 
\begin{eqnarray}
\Rightarrow m &\lsim&  10^3 x / M_{\rm Pl}^3~.
\end{eqnarray}
For the fixed and varying lattices, the above constraint reads
\begin{eqnarray}
\label{constraint_on_m}
 m &\lsim&  70 (e^{-2N_{\rm cl}}) ~M_{\rm Pl} \\ \nonumber
\mbox {and}~~ m &\lsim& 10 ~M_{\rm Pl}~,
\end{eqnarray}
respectively.  Clearly, for any significant proportion of inflation
taking place in the classical era,  Eq.~(\ref{constraint_on_m}a) is
a very strong, and therefore fine-tuned, constraint on the mass of the
inflaton field.  The condition on the inflaton mass becomes however
natural, Eq.~(\ref{constraint_on_m}b), once lattice refinement is
taken into account. 

Equations (\ref{constraint_on_m}) can be compared to restrictions on 
the inflaton mass coming from the WMAP data~\cite{Ringeval:2005yn}, 
for the same inflationary model as the one we consider here, namely:
\be
\label{eq:WMAPdata} 
-5.24 < \log \left( \sqrt{\kappa} \frac{m}{ M_{\rm Pl}} \right) < -5.18~.
\ee 
The fixed lattice case, Eq.~(\ref{constraint_on_m}a) is 
consistent with Eq.~(\ref{eq:WMAPdata}) only if $N_{\rm
cl}< 8.96$, which would suggest that loop quantum gravity effects
should be observable in the spectrum of the CMB temperature anisotropies.
In the lattice refinement case however no such incompatibility between
theoretical and observational constraint exists.

In conclusion, lattice refinement is required to naturally obtain
a successful inflationary model. In contrast to this the fixed lattice
case requires the number of efoldings of inflation that can be
considered classical to be severely fine-tuned to match the
observational data, leading to the conclusion that most of the
inflationary era requires loop quantum gravity corrections to
be included. This is at odds with the startling agreement
between classical CMB calculations and observation.

Finally, it is important to note that whilst
$\tilde{\mu}=\mu_0\mu^{-1/2}$ as been suggested as a natural choice
for the scaling behaviour of the
lattice~\cite{Ashtekar:2006uz2,Vandersloot PhD}, there is no rigorous
justification for this choice. A different choice would result in a
different Hamiltonian constraint equation,
Eq.~(\ref{eq:const_refine}), and all the subsequent analysis would
change.  Dynamic lattice refinement represents a large freedom in the
formulation of LQC that, at present, is fixed only by phenomenological
arguments.  However, deriving the form of this refinement from the
exact LQG equations remains, at present, not possible.

At last, but not least, we would like to bring to the attention of the
reader that an alternative approach to the one we followed here, is to
perform a stability analysis of the dynamical difference equations, as
suggested in Ref.~\cite{Bojowald:2007ra}. More precisely, to
investigating the scale at which the lattice is unable to support the
continuum oscillations, one could perform a von Neumann analysis of
the difference equations on the grid~\cite{0607044}.  Unfortunately,
applying this method to a refining lattice case is more complicated
since plane waves could no longer be used.

\vskip.05truecm 
\acknowledgements 
It is a pleasure to thank Christophe Ringeval for bringing to our
attention Ref.~\cite{Ringeval:2005yn}.  
This work is partially supported by the European Union through the
Marie Curie Research and Training Network \emph{UniverseNet}
(MRTN-CT-2006-035863).


\begin{thebibliography}{10}

\bibitem{Bojowald:2006da}
  M.~Bojowald,
  %``Loop quantum cosmology,''
  Living Rev.\ Rel.\  {\bf 8} (2005) 11
  [arXiv:gr-qc/0601085].
  %%CITATION = 00222,8,11;%%

%\cite{Bojowald:2001xe}
\bibitem{Bojowald:2001xe}
  M.~Bojowald,
  %``Absence of singularity in loop quantum cosmology,''
  Phys.\ Rev.\ Lett.\  {\bf 86}, 5227 (2001)
  [arXiv:gr-qc/0102069].
  %%CITATION = PRLTA,86,5227;%%

%\cite{Ashtekar:2000eq}
\bibitem{Ashtekar:2000eq}
  A.~Ashtekar, J.~C.~Baez and K.~Krasnov,
  %``Quantum geometry of isolated horizons and black hole entropy,''
  Adv.\ Theor.\ Math.\ Phys.\  {\bf 4}, 1 (2000)
  [arXiv:gr-qc/0005126].
  %%CITATION = 00203,4,1;%%

%\cite{Ashtekar:2006uz}
\bibitem{Ashtekar:2006uz}
  A.~Ashtekar, T.~Pawlowski and P.~Singh,
  %``Quantum nature of the big bang: An analytical and numerical 
  % investigation.''
  Phys.\ Rev.\  D {\bf 73}, 124038 (2006)
  [arXiv:gr-qc/0604013].
  %%CITATION = PHRVA,D73,124038;%%

%\cite{Ashtekar:2006uz}
\bibitem{Ashtekar:2006uz2}
  A.~Ashtekar, T.~Pawlowski and P.~Singh,
  %``Quantum nature of the big bang: Improved dynamics.
  %I,''
  Phys.\ Rev.\  D {\bf 74}, 084003 (2006)
  [arXiv:gr-qc/0607039].
  %%CITATION = PHRVA,D73,124038;%%


%\cite{Gibbons:2006pa}
\bibitem{Gibbons:2006pa}
  G.~W.~Gibbons and N.~Turok,
  %``The measure problem in cosmology,''
  arXiv:hep-th/0609095.
  %%CITATION = HEP-TH/0609095;%%

%\cite{Germani:2007rt}
\bibitem{Germani:2007rt}
  C.~Germani, W.~Nelson and M.~Sakellariadou,
  %``On the onset of inflation in loop quantum cosmology,''
  arXiv:gr-qc/0701172.
  %%CITATION = GR-QC/0701172;%%

%\cite{Vandersloot PhD}
\bibitem{Vandersloot PhD}
K.~Vandersloot,
``Loop quantum cosmology'',
Ph.D. Thesis, Pennsylvania State University (2006).

%\cite{Bojowald:2007ra}
\bibitem{Bojowald:2007ra}
  M.~Bojowald, D.~Cartin and G.~Khanna,
  %``Lattice refining loop quantum cosmology, anisotropic models and
  %stability,''
  arXiv:gr-qc/0704.1137.
  %%CITATION = ARXIV:0704.1137;%%

%\cite{Ashtekar:2003hd}
\bibitem{Ashtekar:2003hd}
  A.~Ashtekar, M.~Bojowald and J.~Lewandowski,
  %``Mathematical structure of loop quantum cosmology,''
  Adv.\ Theor.\ Math.\ Phys.\  {\bf 7}, 233 (2003)
  [arXiv:gr-qc/0304074].
  %%CITATION = 00203,7,233;%%

%\cite{Bojowald:2002ny}
\bibitem{Bojowald:2002ny}
  M.~Bojowald,
  %``Quantization ambiguities in isotropic quantum geometry,''
  Class.\ Quant.\ Grav.\  {\bf 19}, 5113 (2002)
  [arXiv:gr-qc/0206053].
  %%CITATION = CQGRD,19,5113;%%

%\cite{Vandersloot:2005kh}
\bibitem{Vandersloot:2005kh}
  K.~Vandersloot,
  %``On the Hamiltonian constraint of loop quantum cosmology,''
  Phys.\ Rev.\  D {\bf 71}, 103506 (2005)
  [arXiv:gr-qc/0502082].
  %%CITATION = PHRVA,D71,103506;%%

%\cite{Vandersloot:2005}
\bibitem{Vandersloot:2005}
  K.~Vandersloot,
  %``Loop quantum cosmology and the $k=-1$ Robertson-Walker model''
  Phys.\ Rev.\  D {\bf 75}, 023523 (2007)
  [arXiv:gr-qc/0612070].


%\cite{Vandersloot:2005}
\bibitem{Vandersloot:2005_2}
  K.~Vandersloot,
  %``Loop quantum cosmology of the $k=+1$ FRW model''
  Phys.\ Rev.\  D {\bf 75}, 023523 (2007)
  [arXiv:gr-qc/0612070].


%\cite{Ashtekar:2007}
\bibitem{Ashtekar:2007}
  A.~Ashtekar, T.~Pawlowski, P.~Singh, K.~Vandersloot,
  %``Loop quantum cosmology of the $k=+1$ FRW model''
  Phys.\ Rev.\  D {\bf 75}, 023523 (2007)
  [arXiv:gr-qc/0612104].

%\cite{Bojowald:2002gz}
\bibitem{Bojowald:2002gz}
  M.~Bojowald,
  %``Isotropic loop quantum cosmology,''
  Class.\ Quant.\ Grav.\  {\bf 19}, 2717 (2002)
  [arXiv:gr-qc/0202077].
  %%CITATION = CQGRD,19,2717;%%

%\cite{Bojowald:2004xq}
\bibitem{Bojowald:2004xq}
  M.~Bojowald, J.~E.~Lidsey, D.~J.~Mulryne, P.~Singh and R.~Tavakol,
  %``Inflationary cosmology and quantization ambiguities in semi-classical  
  %loop quantum gravity,''
  Phys.\ Rev.\  D {\bf 70}, 043530 (2004)
  [arXiv:gr-qc/0403106].
  %%CITATION = PHRVA,D70,043530;%%

%\cite{cobe}
\bibitem{cobe}
  C.~L.~Bennett, {\sl et. al}, Astrophy.\ J.\ {\bf 464}, 1 (1996)
  [arXiv:astro-ph/9601067].

%\cite{Ringeval:2005yn}
\bibitem{Ringeval:2005yn}
  C.~Ringeval, P.~Brax, C.~van de Bruck and A.~C.~Davis,
  %``Boundary inflation and the WMAP data,''
  Phys.\ Rev.\  D {\bf 73} (2006) 064035
  [arXiv:astro-ph/0509727].
  %%CITATION = PHRVA,D73,064035;%%


%\cite{0607044}
\bibitem{0607044}
 J.~Rosen, J.-H.~ Jung and G.~Khanna,
 Class.\ Quant.\ Grav.\  {\bf 23}, 7075 (2006)
 [arXiv:gr-qc/0607044].


\end{thebibliography}
\end{document}